\documentclass{optica-article}

\journal{opticajournal} 

\articletype{Research Article}

\usepackage{lineno}

\begin{document}

\title{
Nonlocal Manipulation of Backflow with Quantum Correlations
}

\author{Ya Xiao,\authormark{1,$\dag$,*} Zhen-Fei Zhang,\authormark{1,$\dag$} Yan-Xin Rong,\authormark{1} Kai Sun,\authormark{2,3} Jin-Shi Xu,\authormark{2,3} and Yong-Jian Gu\authormark{1,4,5,*}}
 \address{\authormark{1}College of Physics and Optoelectronic Engineering, Ocean University of China, Qingdao 266100, China.\\
 \authormark{2}Laboratory of Quantum Information, University of Science and Technology of China, Hefei 230026, China.\\
\authormark{3}Anhui Province Key Laboratory of Quantum Network, University of Science and Technology of China, Hefei, Anhui 230026, China.\\
 \authormark{4}Engineering Research Center of Advanced Marine Physical Instruments and Equipment (Ministry of Education), Ocean University of China, Qingdao 266100, China.\\
\authormark{5}Qingdao Key Laboratory of Optics and Optoelectronics, Ocean University of China, Qingdao 266100, China.\\

 \authormark{\dag}The authors contributed equally to this work.\\}

\email{\authormark{*}xiaoya@ouc.edu.cn\\
\authormark{*}yjgu@ouc.edu.cn}


\begin{abstract*} 

Quantum correlations are central resources for quantum information processing, yet their ability to manipulate dynamical transmission processes remains largely unexplored. Here, we investigate this ability through backflow, a uniquely interference phenomenon in which local probability flow propagates opposite to the momentum direction. We report the first nonlocal manipulation of backflow in double-slit interference using polarization-path-entangled photons. By performing local measurements on one photon, we remotely engineer the relative amplitude and phase of the two paths associated with its partner, manipulating the emergence, spatial distribution, and propagation dynamics of backflow without directly accessing the interfering system. Combining weak measurements to extract the transverse momentum and reconstruct Bohmian trajectories, we provide a direct visualization of the manipulation process with single-pixel spatial resolution. Furthermore, using Werner states with tunable correlation strengths, we reveal a distance-dependent resource requirement for nonlocal backflow manipulation: the minimum correlation strength required to induce backflow increases with propagation distance, progressing from entanglement to EPR-steering and ultimately Bell nonlocality. Our results show quantum correlations as operational resources for manipulating transmission dynamics and open new avenues for non-contact manipulation of fragile or inaccessible systems.

\end{abstract*}

\section{Introduction}
Quantum correlations are widely recognized as fundamental resources for quantum information processing, enabling tasks ranging from secure communication \cite{BENNETT20147} and computational acceleration \cite{calculation} to enhanced measurement precision \cite{science.1104149}. While their role in information-processing tasks is well established, a broader question remains largely unexplored: can quantum correlations serve as operational resources for manipulating dynamical transmission processes? Addressing this question would extend quantum resource theories beyond information tasks toward the active manipulating of probability current, energy flow and wave dynamics, opening new possibilities for non-contact manipulate of microscopic systems.

Quantum backflow provides an ideal platform for addressing above question. It is a counterintuitive interference phenomenon in which a particle whose momentum spectrum is strictly positive (or negative) nonetheless exhibits a local probability current directed opposite to its momentum. Because backflow is highly sensitive to the relative amplitudes and phases of interfering modes, it offers a unique testbed for investigating whether quantum correlations can be exploited to manipulate transmission dynamics. Beyond its fundamental significance, backflow has attracted increasing attention because of its connections to quantum arrival times \cite{Allcock1969-1,Allcock1969-2,Allcock1969-3} and tunneling dynamics \cite{trillo2023}, and emerging applications in nanoparticle manipulation \cite{tian2021,xie2025optical}, enhanced chiral molecular responses \cite{tang2011}, super-resolution imaging~\cite{rogers2012super,pu2021unlabeled}, and high-sensitivity optical metrology~\cite{yuan2019detecting}.

First discovery by Allcock \cite{Allcock1969-1,Allcock1969-2,Allcock1969-3}, quantum backflow has been extensively theoretically investigated in a wide range of scenarios, including nonrelativistic free particles \cite{Bracken1994}, systems with spin–orbit interactions \cite{mardonov2014}, linear potentials \cite{melloy1998}, noisy environments \cite{albarelli2016,bostelmann2017,goussev2019,vandijk2019}, and even relativistic regimes \cite{melloy1998probability,ashfaque2019relativistic}. 

Experimentally, backflow-like effects have been observed in classical optical systems using sub-oscillatory wavefunctions \cite{eliezer2020}, plane wave interference \cite{daniel2022}, or vortex beam interference \cite{ghosh2023}. Recently, we reported the first experimental realization of azimuthal backflow in a  single-photon system \cite{zhang2025}. Despite these advances, existing studies have focused almost exclusively on the observation  of backflow.

However, for both fundamental science and practical applications, observation alone is insufficient. The ability to actively manipulate transmission processes is essential for transforming quantum phenomena into useful functionalities. In the context of backflow, a central open question is whether the phenomenon can be manipulated nonlocally through quantum correlations, thereby enabling manipulation without direct access to the transmission system. More important is determining what type and strength of quantum correlations are required to sustain such manipulation. Resolving these questions would establish a direct operational connection between quantum resources and backflow, while providing new strategies for manipulating fragile or  inaccessible systems.

Here, we theoretically and experimentally demonstrate that quantum correlations enable nonlocal manipulation of backflow. Using entangled photon pairs, one photon traverses a birefringent double-slit apparatus where backflow is investigated, while the other is subjected to local projective measurements that remotely engineer the relative amplitude and phase between the two slits. By employing weak measurement techniques, we directly quantify this manipulation through the measurement of probability currents and the reconstruction of Bohmian trajectories. The results show that adjusting the relative amplitude promotes the occurrence of backflow near the slit with lower amplitude, whereas varying the relative phase induces a phase-dependent symmetry breaking of the backflow distribution. Furthermore, time-programmed modulation of the relative phase enables the particle to remain within the backflow region during propagation, potentially extending the operational range of backflow-based manipulation schemes. Most importantly, by employing Werner states with tunable correlation strengths, we uncover a distance‑dependent resource requirement: the minimum correlation strength needed to nonlocally induce backflow increases with propagation distance, progressing from entanglement to EPR-steering and ultimately to Bell nonlocality. These findings establish a quantitative link between quantum correlations and backflow, transforming backflow from a passive quantum effect into an actively controllable process and opening new avenues for developing non‑contact manipulation techniques in photonic systems.

\section{Protocol for nonlocal manipulation of backflow}
To describe the nonlocal manipulation of backflow, we consider an arbitrary two-qubit state $\rho_{AB}$ shared between photon A, encoded in polarization, and photon B, encoded in the path degree of freedom after a double slit. The spatial wave functions of the upper and lower slits are $\psi_{u}(x,z) \equiv g(x-d/2,z)$ and $\psi_{l}(x,z)\equiv  g(x+d/2,z)$, where $d$ is the slit separation,  $x$ and $z$ denote the transverse and propagation coordinates, respectively. The normalized Gaussian wavepacket is
\begin{equation} 
g(x\pm d/2,z)=\left(\frac{2}{\pi \omega_z^{2}}\right)^{\frac{1}{4}} 
\exp\!\left[-\frac{(x\pm d/2)^{2}}{\omega_z^{2}} + i\Bigl(kz - \arctan\!\bigl(\tfrac{z}{z_R}\bigr) + \frac{k (x\pm d/2)^{2}}{2 R_z}\Bigr)\right]. 
\label{Gaussian wavepacket} 
\end{equation}
Here, $\omega_z $, ${z_R} $ and $k $ denote the beam radius, Rayleigh range, and wavenumber~\cite{zhang2025}, respectively.

Photon A is projected onto the polarization basis
$\left|m\right\rangle_A=\cos\theta\left|H\right\rangle_A - e^{-i\varphi}\sin\theta\left|V\right\rangle_A$,
 where $\theta\in[0,\pi/2]$ and $\varphi\in[0,2\pi]$. This local projection remotely prepares photon B in the conditional state

\begin{equation}
\rho_{B|m} = \frac{\operatorname{Tr}_A\big[(|m\rangle_A\langle m| \otimes I_B)\,\rho_{AB}\big]}{\operatorname{Tr}\big[(|m\rangle_A\langle m| \otimes I_B)\,\rho_{AB}\big]},
\label{stateB}
\end{equation} 
thereby manipulating the relative amplitude and phase between the two paths of photon B without directly accessing the interfering system. Consequently, the interference responsible for backflow can be manipulated through local measurements on photon A.

Expressing the conditional state as a convex mixture of pure states, $\rho_{B|m}=\sum_n w_n|\psi_n\rangle\langle\psi_n|$, the probability current is given by~\cite{luis2015dynamics}
\begin{equation}
J_x^{\rho_{B|m}}(x,z) = \sum_n w_n \, J_x^{n}(x,z) = \sum_n w_n\, \frac{1}{m}\, \rho_x^{n}(x,z)  k_x^{n}(x,z),
\label{eq:current}
\end{equation}
where $w_n \ge 0$, $\sum_n w_n = 1$, and $J_x^{n}(x,z)$ is the probability current of each pure component.

\begin{figure}[!hbtp]
	\centering
	\includegraphics[width=0.85\linewidth]{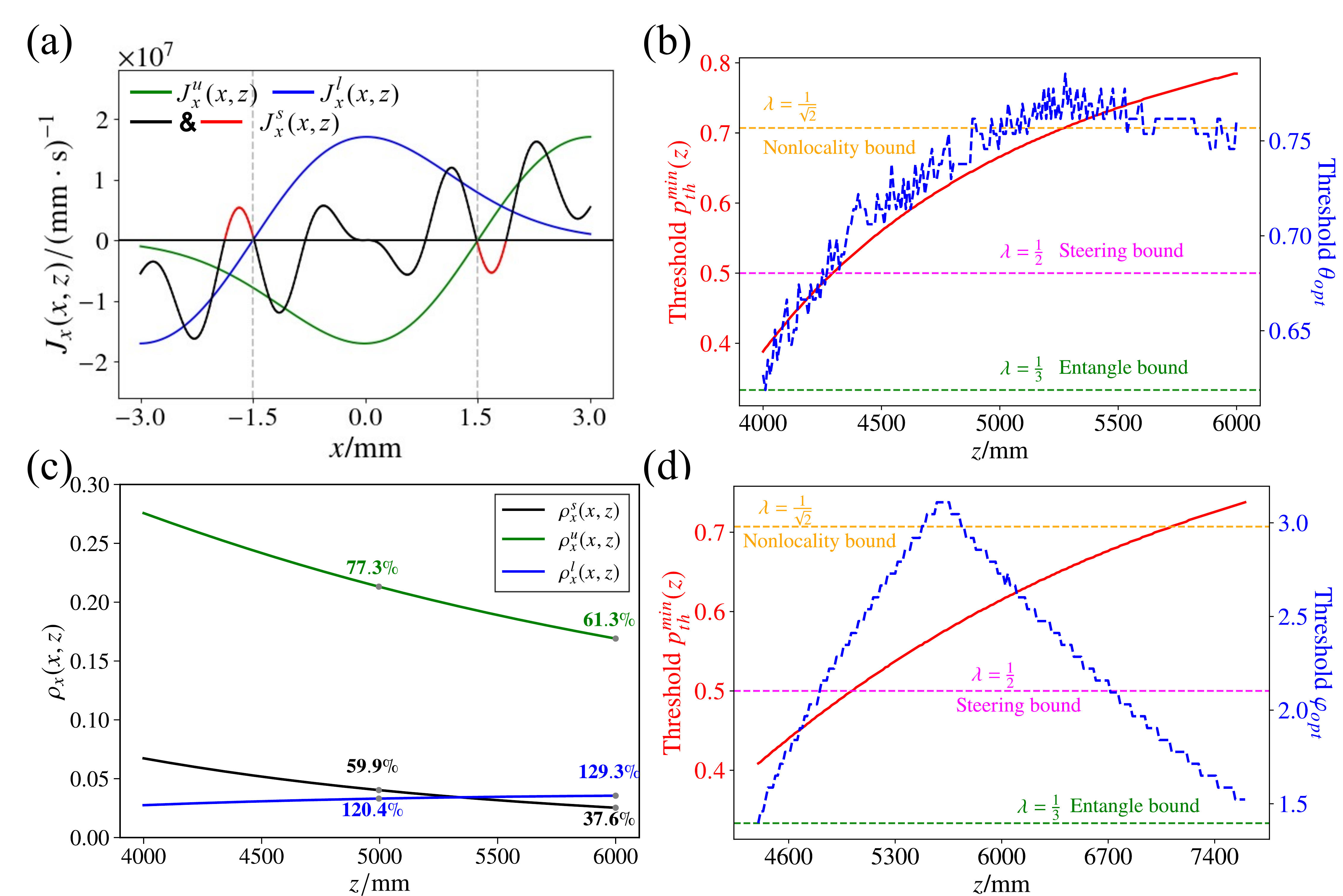}
	\caption{Manipulate backflow in double-slit interference. 
      (a) Probability currents $J_{x}^{s}(x,z)$, $J_{x}^{u}(x,z)$, and $J_{x}^{l}(x,z)$ at $z = 4200~\text{mm}$ for the maximally entangled state $|\psi_s\rangle$ with photon A projected onto $(|H\rangle_A - |V\rangle_A)/\sqrt{2}$. Red curves indicate backflow regions.
    (b) Minimum quantum correlation strength and optimal measurement angle versus propagation distance for the amplitude modulation protocol.
    (c) Probability densities $J_{x}^{s}(x,z)$, $J_{x}^{u}(x,z)$, and $J_{x}^{l}(x,z)$ as function of propagation distance for the amplitude modulation protocol. The inset  values normalized to their respective initial values.
    (d) Minimum quantum correlation strength and optimal measurement angle versus propagation distance for the phase modulation protocol. Other parameters (also used below): $\lambda=810 \mathrm{nm}$, $\omega_o=0.36 \mathrm{mm}$, $d=3 \mathrm{mm}$.}
	\label{theory}
\end{figure}

For an individual slit, spatial wave functions diverge outward, the probability current is always positive for $x > d/2$ and negative for $x < -d/2$. Backflow occurs when coherent interference reverses this outward flow, namely when 
$J_x^{\rho_{B|m}}(x,z) < 0$ for $x > d/2$, or equivalently $J_x^{\rho_{B|m}}(x,z) > 0$ for $x < -d/2$, as illustrated by the red curves in Fig.~\ref{theory}(a). Since the interference is dynamically determined by the remotely engineered relative amplitude and phase, its emergence, spatial distribution, and propagation dynamics can all be manipulated nonlocally.

To quantify the quantum resources required for nonlocal backflow manipulation, we consider photon A and photon B initially shared a polarization-path Werner state  $\rho_{AB}^W = p\,|\psi_s\rangle\langle\psi_s| + \frac{1-p}{4}\,I_A \otimes I_B^{\mathrm{path}}$,
where $|\psi_s\rangle=
(|H\psi_u(x,z)\rangle-|V\psi_l(x,z)\rangle)/\sqrt{2}$ and $p$ is the purity parameter. The Werner states are entangled for $p>1/3$, steerable for $p>1/2$, and Bell nonlocal for $p>1/\sqrt2$ \cite{werner1989quantum,Wiseman98,acin2006grothendieck}. After projection, the conditional state consists of a coherent interference component weighted by $p$ and an incoherent background weighted by $1-p$. Consequently, its probability current can be decomposed as $J_x^{W}(x,z)= p\,J_x^{s}(x,z) + \frac{1-p}{2} J_x^{u}(x,z) + \frac{1-p}{2} J_x^{l}(x,z).$ Because the incoherent component contributes only the outward single-slit transverse probability currents, decreasing $p$ progressively suppresses the interference-induced reversal of the transverse probability current and eventually eliminates backflow.

To characterize this behavior quantitatively, we define the total backflow length $f_{\rm length}(z,\theta,\varphi,p)$ $= \int_{d/2}^{\infty} \Theta[-J_x^{\rho_{B|m}}(x,z)] \, \rm{d}x + \int_{-\infty}^{-d/2} \Theta[J_x^{\rho_{B|m}}(x,z)] \, \rm{d}x$ as the total spatial extent over which the probability current reverses direction, where $\Theta[ \dot  ]$ is the Heaviside step function~\cite{kanwal1998generalized}.  As shown in Appendix A, for fixed $(z, \theta,\varphi)$,  the total backflow length increases monotonically with $p$, implying that each propagation plane possesses a unique critical purity $p_{\rm th}(z, \theta,\varphi)$ above which the backflow occurs. Optimizing projection angles yields $ p_{\rm th}^{\rm min}(z)=\rm{min}_{\theta,\varphi}$$ p_{\rm th}(z,\theta,\varphi)$, which defines the minimum quantum correlation required for nonlocal backflow manipulation at propagation distance $z$.

Figure~\ref{theory}(b) shows $p_{\rm th}^{\rm min}(z)$ for the amplitude-modulation protocol (fixed $\varphi$ and varying $\theta$), which increases monotonically with propagation distance. This behavior originates from free-space divergence. Specifically, as the wavepackets propagate, the transverse spreading causes the probability density of the interference wave function to decay more rapidly than that of the single-slit wave function.  This differential decay impedes the ability of the interference-induced backward current to surpass the forward single-slit current, thereby suppressing backflow (see Fig. 1(c)). Since $J_x^{W}(x,z)$ is directly determined by the purity parameter $p$, the minimum quantum correlation required for nonlocal backflow manipulation also increases with propagation distance. Remarkably, the monotonically increasing sequentially crosses the entanglement, EPR-steering, and Bell-nonlocality boundaries of Werner states, revealing a distance-dependent quantum resource requirement for nonlocal backflow manipulation. The same analysis applies to the phase-modulation protocol, i.e., fixed $\theta$ and manipulating $\varphi$, yielding the analogous results when $\theta=\pi/4$ shown in Fig.~\ref{theory}(d).

\section{Experimental nonlocal manipulation of backflow}
\begin{figure}[!hbtp]
	\centering
	\includegraphics[width=0.8\linewidth]{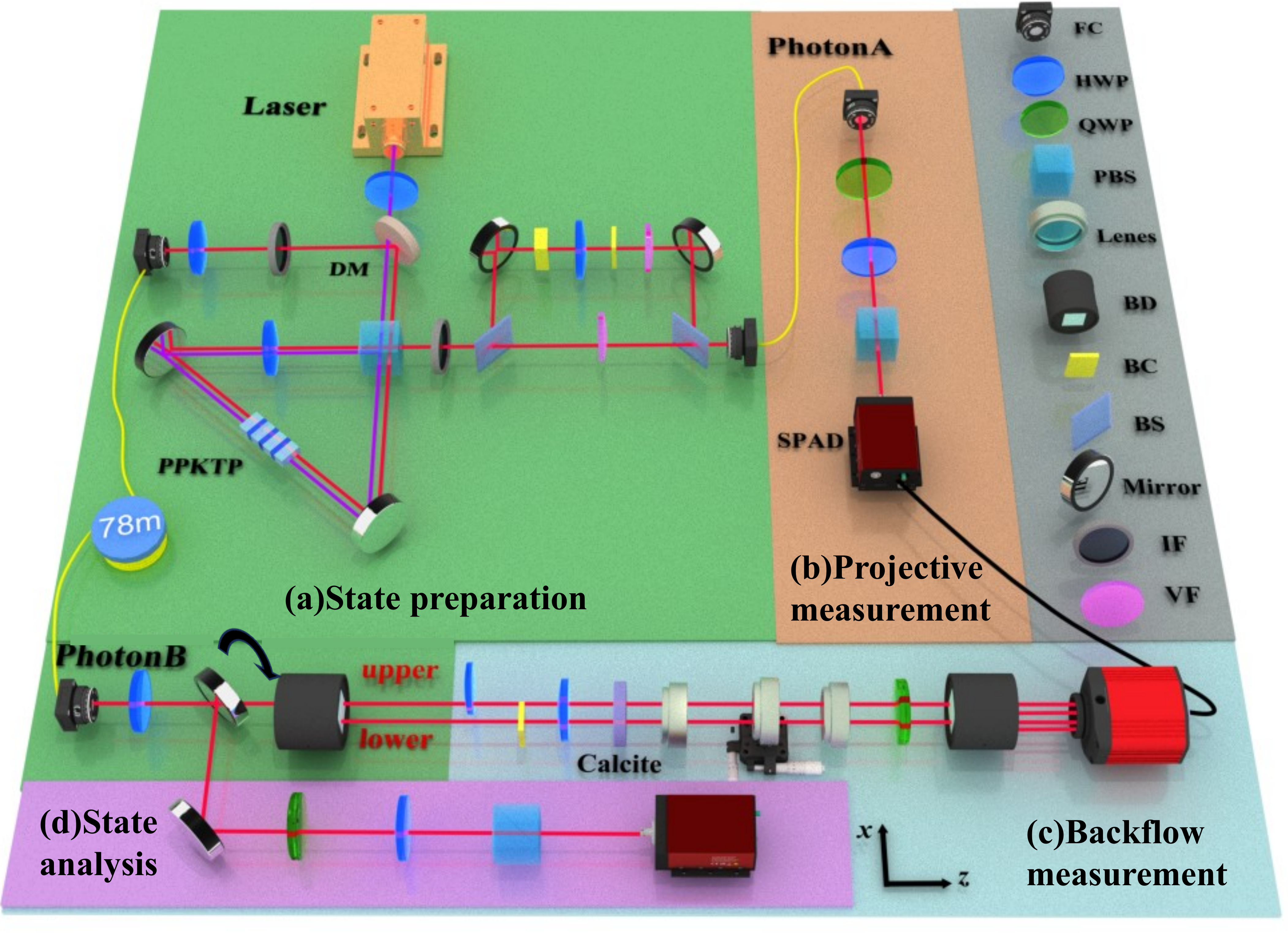}
	\caption{Experimental setup. (a) Source of Werner‑state entangled photon pair A and B. (b) Projective measurement system for photon A. (c) Backflow detection system for photon B. (d) Correlation‑strength analysis of the prepared state. HWP:Half-wave plates; QWP: Quarter-wave plates; DM: Dichroic mirror; PBS: Polarizing beam splitters; VOA: Variable Optical Attenulator; BD: Beam displacers; L1-L3: Lens; BC: Birefringent crystal; PPKTP: periodically poled KTiOPO$_4$ crystal; SPAD: Single photon avalanche diode; ICCD: Intensified CCD.}
	\label{setup}
\end{figure}

The experimental setup is illustrated in Fig.~\ref{setup}. 
A diagonally polarized 405\,nm laser is directed into a Sagnac interferometer to bidirectionally pump a type-II periodically poled KTiOPO$_4$ (PPKTP) crystal, generating maximally polarization-entangled photon pairs in the state $(|HV\rangle - |VH\rangle)/\sqrt{2}$. Polarization‑maintained by two half‑wave plates (HWPs), photon B then enters a beam displacer (BD) that transmits $|V\rangle$ and displaces $|H\rangle$ laterally by 3\,mm, preparing upper-wave function $\psi_u(x,z)$ for  $|V\rangle$ and lower-slit wave function $\psi_l(x,z)$  for $|H\rangle$, respectively. Photon A is directed into an unbalanced interferometer, where a beam splitter (BS) divides it into a transmitted and a reflected path. The quantum state in the transmitted path remains unchanged. In the reflected path, two sufficiently long birefringent crystals (BCs) separated by a HWP at $22.5^\circ$ completely destroy the coherence between $|H\rangle$ and $|V\rangle$. Recombining the paths prepares the required Werner state $\rho_{AB}^W$ defined in the previous section. Two variable filters (VFs) are used to adjust the purity parameter $p$.  

Followed by an HWP for polarization alignment and a BC for optical path compensation. The transverse probability current of photon B is determined by the probability density and the weak value of its transverse momentum~\cite{xiao2017}. To obtain \(\langle k^{j}_x(x_m^{z_n},z_n) \rangle_w\) at position  $ x_m^{z_n} $ and  $z_n$ plane, photon B is prepared in the diagonal polarization state, $ (|H\rangle + |V\rangle)/\sqrt{2} $, using HWP5 and then normally incident on a 0.579-mm thick calcite crystal with its optical axis oriented at \(42^\circ\) in the \(x\)-\(z\) plane relative to the \(z\) axis. This interaction induces a transverse momentum-dependent polarization rotation, $|H\rangle + |V\rangle \rightarrow |H\rangle + e^{i\zeta k^{j}_x(x_m^{z_n},z_n)/k}|V\rangle$, with a coupling strength \(\zeta = 413\)~\cite{zhang2025}. The resulting rotation is measured on the right-hand/left-hand circular basis $|R/L\rangle= (|H\rangle \pm i|V\rangle)/\sqrt{2}$, using a QWP oriented at \(45^\circ\) and a BD. A three-lens imaging system (L1-L3), with the middle lens (L2) translatable along the $z$ axis, enables measurements over propagation distances ranging from the near field to the far field. Photon counts \(N^{j}_{R}(x_m^{z_n},z_n)\) and \(N^{j}_{L}(x_m^{z_n},z_n)\), corresponding to right-hand and left-hand circular polarization, are recorded by an ICCD camera. The weak value of the transverse momentum is then extracted as
\begin{equation}
\langle k^{j}_x(x_m^{z_n},z_n) \rangle_w = \frac{k}{\xi}\,\arcsin\!\left[\frac{N^{j}_R(x_m^{z_n},z_n) - N^{j}_L(x_m^{z_n},z_n)}{N^{j}_R(x_m^{z_n},z_n) + N^{j}_L(x_m^{z_n},z_n)}\right],
\end{equation}
and the probability current $\langle J^{j}_x(x_m^{z_n},z_n) \rangle_w$ can be expressed as 
\begin{equation}
\langle J^{j}_x(x_m^{z_n},z_n) \rangle_w =\frac{\lambda c }{2 \pi \hbar}\rho_{j}(x_m^{z_n},z_n)\langle k^j_x(x_m^{z_n},z_n)\rangle_w,
\end{equation}
where $\rho_{j}(x_m^{z_n},z_n)=\frac{N^{j}_R(x_m^{z_n},z_n) + N^{j}_L(x_m^{z_n},z_n)}{\sum_{m}(N^{j}_R(x_m^{z_n},z_n) + N^{j}_L(x_m^{z_n},z_n))}$ is the probability density at position $ x_m^{z_n} $ and  $z_n$ plane. To gain a deeper understanding of the dynamics of backflow, we can reconstruct the Bohmian trajectories $T_{j}(x_m^{z_n},z_n)$ using the iterative relation $x_m^{z_{n+1}} = x_m^{z_{n}} +(z_{n+1} - z_n) v_x(x_m^{z_n},z_n)/\sqrt{c^{2}- v_x^2(x_m^{z_n},z_n)}$ with $ v_x(x_m^{z_n},z_n)= k^{j}_x(x_m^{z_n},z_n) \rangle_w/m$ \cite{xiao2017,xiao2019}.

\begin{figure}[!hbtp]
	\centering
	\includegraphics[width=0.85\linewidth]{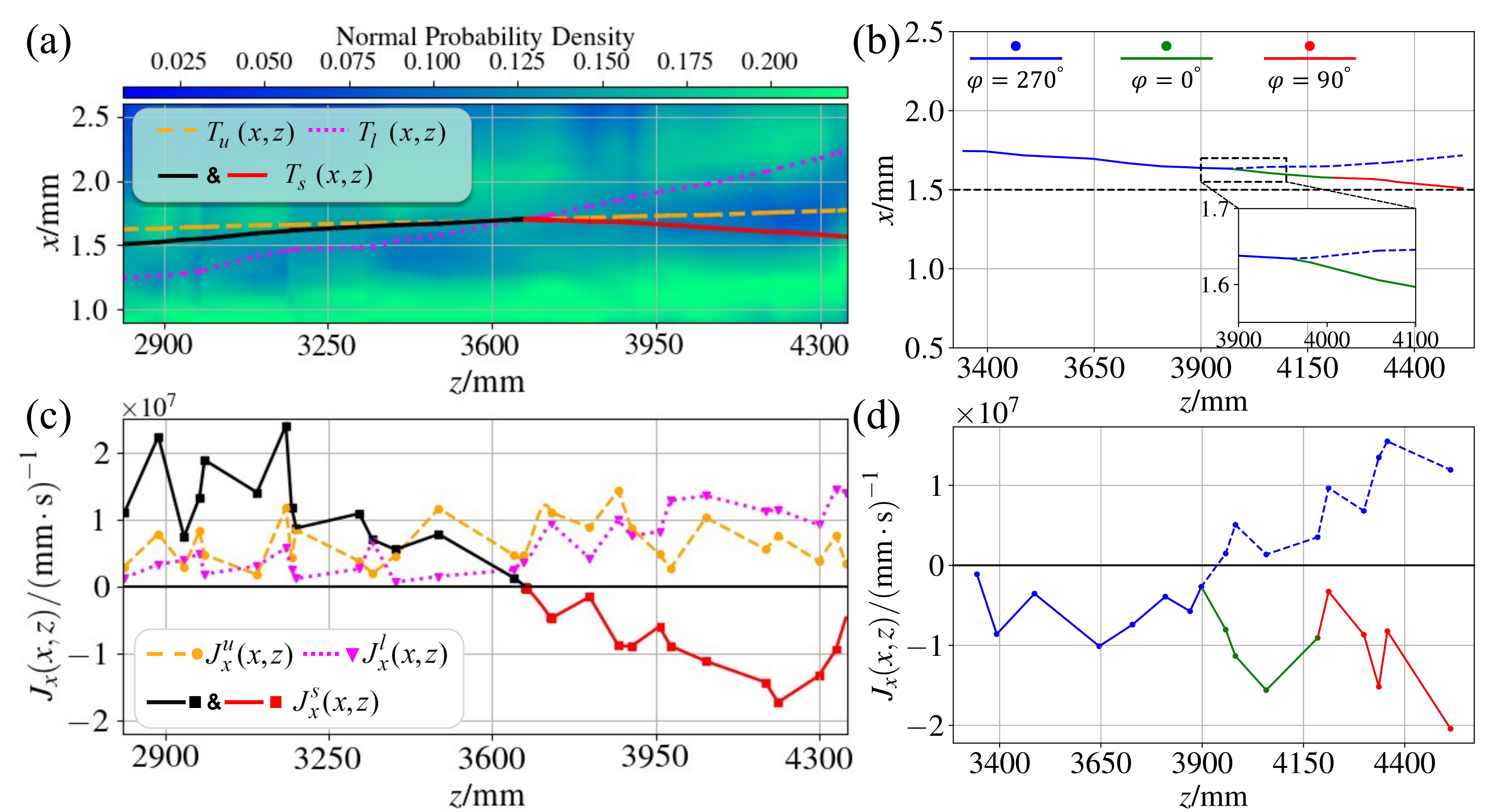}
	\caption{Dynamic manipulation of backflow in double‑slit interference. (a) Bohmian trajectories guided respectively by: the upper-slit wavefunction $\psi_{u}(x,z) $, beginning at $x=1.625$ mm; the lower-slit wavefunction $\psi_{l}(x,z) $, beginning at $x=1.244$ mm; and the double-slit superposition wavefunction $\psi_{s}(x,z)=(\psi_{u}(x,z)+\psi_{l}(x,z) )/\sqrt{2}  $, beginning at $x=1.508$ mm. Backflow regions in (a) is marked by red curve. (b) Bohmian trajectories of photon B originate from the position $(x_0=1.74\ \mathrm{mm}, z_0=3344\ \mathrm{mm})$, with $\varphi(z)$ dynamically modulated at suitable propagation distances. (c) Evolution of probability current $J^{j}_x(x,z)$ along trajectory $T_{\mathrm{s}}(x,z)$ with propagation distance $z$, where  $j=\{u,l,s\}$. (d) Evolution of probability current along the trajectories shown in (b). Backflow is indicated by a negative probability current in (c) and (d).}
	\label{exp}
\end{figure}

As illustrated in Fig.~\ref{exp}(a), Bohmian trajectories $T_u(x,z)$, $T_l(x,z)$, and $T_s(x,z)$, originating from $x = 1.625$ mm, $x = 1.244$ mm, and $x = 1.508$ mm, respectively, were reconstructed over the range $z = 2813$ mm to $z = 4357$ mm using 26 imaging planes. These trajectories are guided by the upper-slit wavefunction $\psi_u(x,z)$, the lower-slit wavefunction $\psi_l(x,z)$, and their coherent superposition $\psi_s(x,z) = (\psi_u(x,z) + \psi_l(x,z))/\sqrt{2}$ corresponding to the cases where photon A, which shares a maximally entangled state with photon B, is projected onto  $|H\rangle_A $, $|V\rangle_A$ and $(|H\rangle_A - |V\rangle_A)/\sqrt{2}$, respectively. All three trajectories converge at $z = 3667$ mm. Beyond this point, $T_u(x,z)$ and $T_l(x,z)$ continue propagating in the $+x$ direction, whereas $T_s(x,z)$ reverses toward the $-x$ direction,  providing visually intuitive evidence of backflow. The evolution of the probability current $J_x^i(x,z)$ along the superposition trajectory $T_s(x,z)$ is presented in Fig.~\ref{exp}(c). Pronounced sign inversion in $J_x^{\mathrm{s}}(x,z)$, relative to $J_x^{\mathrm{u}}(x,z)$ and $J_x^{\mathrm{l}}(x,z)$, occurs beyond the convergence point, confirming backflow. 

Moreover, the measurement of photon A instantaneously changes the probability current of photon B, inducing a directional shift in its trajectory depending on the measurement outcome. By modulating $\varphi(z)$ at suitable propagation distances, this nonlocal effect enables photon B to remain within the backflow region.Specifically, Fig.~\ref{exp}(b) shows that when photon A performs a static phase measurement with $\varphi=90^\circ$, photon B, starting at an initial position $(x_0=1.74\ \mathrm{mm}, z_0=3344\ \mathrm{mm})$ within the backflow region, exits the backflow region at $z=3870$ mm (blue curves). However, by sequentially modulating photon A's phase to $\varphi(z)=0^\circ$ at $z=3870$ mm and then to $\varphi(z)=270^\circ$ at $z=4185$ mm, photon B is effectively confined within the backflow region throughout its entire propagation. These trajectories are reconstructed from $z = 3344 \ \mathrm{mm}$ to $z = 4513 \ \mathrm{mm}$ using 17 different planes.  As shown in Fig.~\ref{exp}(d), for $x_0= 1.74\ \mathrm{mm} $, the transverse probability current remains negative when $z > 3344$ mm, again confirming the successful confinement of photon B within the backflow region through dynamic phase modulation.

\begin{figure}[!hbtp]
	\centering
	\includegraphics[width=0.85\linewidth]{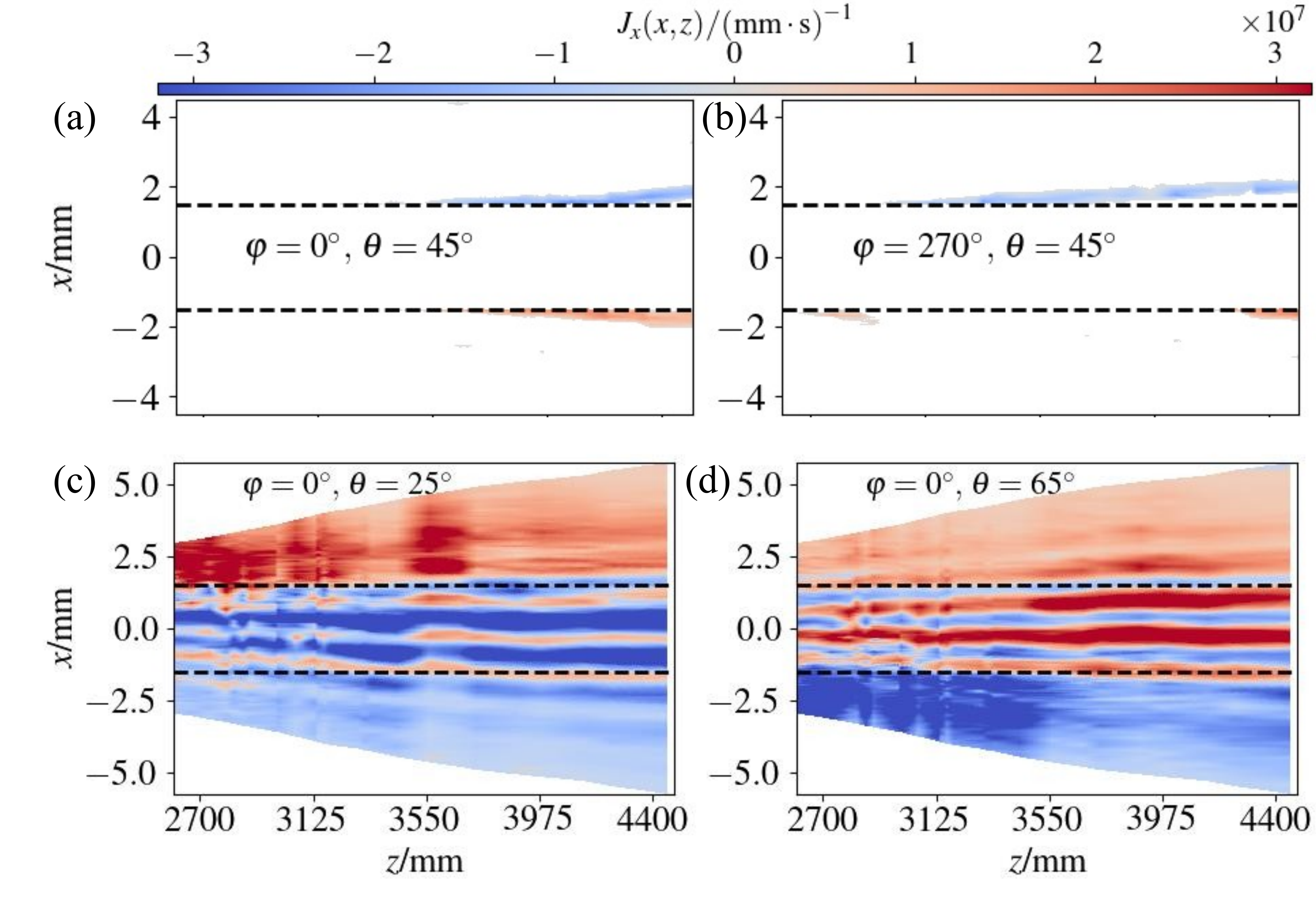}
	\caption{Manipulation of backflow via static phase and amplitude modulation. (a, b) Backflow regions for photon B when photon A is projected onto the state $\left|m\right\rangle_A = (\left|H\right\rangle - e^{-i\varphi} \left|V\right\rangle)/\sqrt{2}$ for $\varphi =  0^\circ$ and $270^\circ$, respectively. (c, d) Probability current distribution of photon B in the $x$-$z$ plane is shown when photon A is projected onto the state $\left|m\right\rangle_A =\cos\theta\left|H\right\rangle - \sin\theta\left|V\right\rangle$ for $\theta=25^\circ$ (c) and $\theta=65^\circ$ (d). The red regions within $x < -d/2$ and blue regions with $x > d/2$ denote the existence of backflow. }
	\label{amplitude}
\end{figure}

Beyond dynamical schemes, we also employ a static phase modulation by post-selecting photon A in the polarization state $\left|\theta\right\rangle_A = (\left|H\right\rangle - e^{-i\varphi} \left|V\right\rangle)/\sqrt{2}$, under which the relative phase $\varphi$ remains invariant during propagation. Figure~\ref{amplitude}(a) and (b) present the spatial distributions of the backflow region for $\varphi= 0^\circ$ and $270^\circ$, respectively (corresponding results for $\varphi= 90^\circ$ and $180^\circ$  are provided in Appendix B). Clearly, the backflow regions are symmetric about $x = 0$ when $\varphi= 0^\circ$ and $180^\circ$, whereas this symmetry is broken at $\varphi = 90^\circ$ and $270^\circ$. The backflow phenomenon primarily occurs near the lower slit when $\varphi \in (0^\circ, 180^\circ)$, whereas it shifts toward the upper slit when $\varphi \in (180^\circ, 360^\circ)$.  

Figure~\ref{amplitude}(c) and (d) illustrate the results of nonlocal manipulation of backflow with static amplitude modulation, implemented by projecting photon A onto the state $\left|\theta\right\rangle_A =\cos\theta\left|H\right\rangle - \sin\theta\left|V\right\rangle$ while varying $\theta$. Clearly, the measurement on photon A changes the probability that photon B passes through the upper and lower slits, thereby modifying the amplitude of the corresponding wavefunctions. It can be observed that in the near-field region, backflow emerges earlier within the destructive interference regions near the slit with a lower amplitude (the lower slit for $\theta=25^\circ$ and the upper slit for $\theta=65^\circ$).  This behavior arises from the requirement that interference visibility exceed a critical threshold for backflow to occur \cite{zhang2025}. 

\begin{figure}[!hbt]
	\centering
	\includegraphics[width=0.8\linewidth]{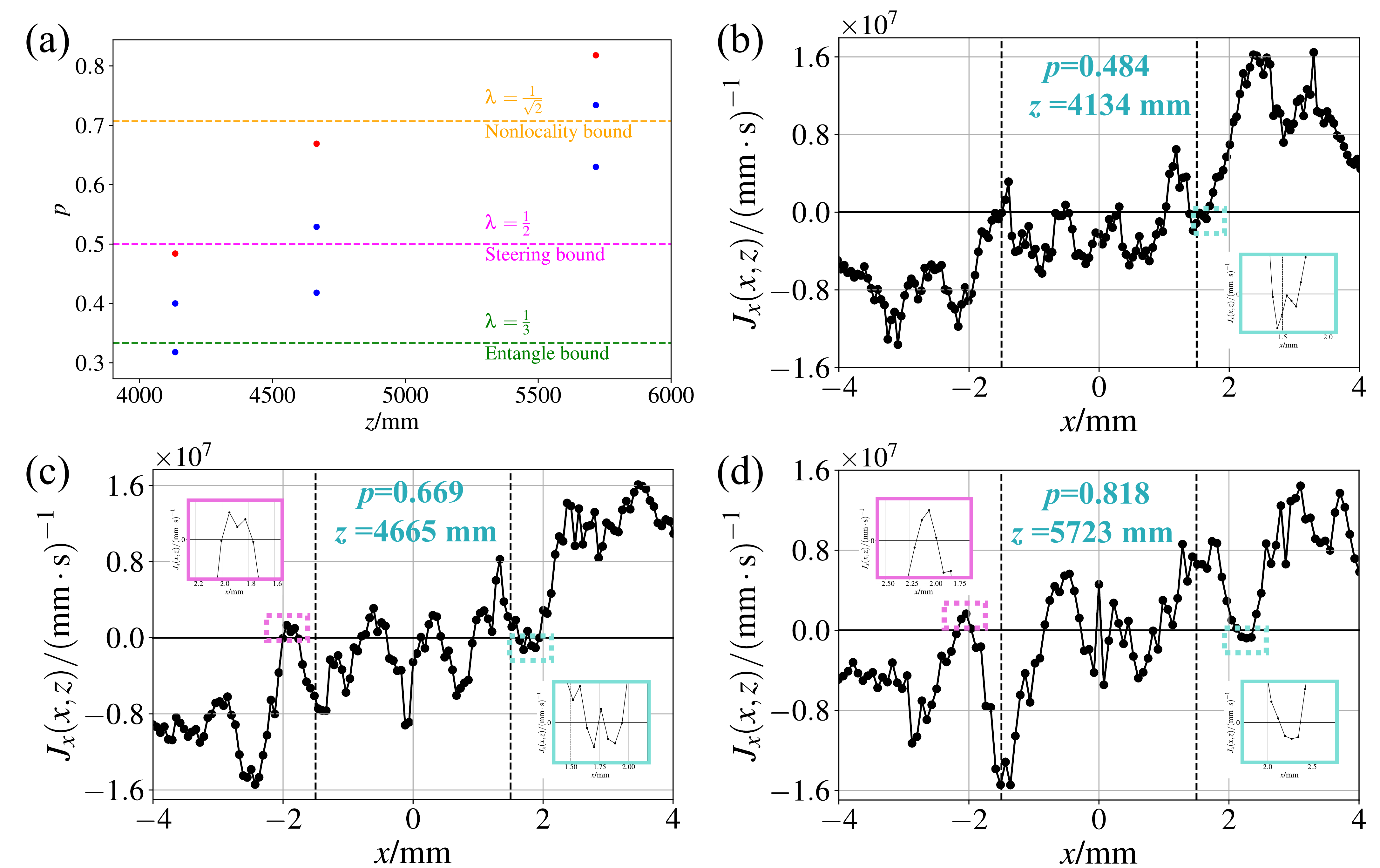}
	\caption{Distance‑dependent quantum resource requirement for nonlocally manipulating the emergence of backflow.
    (a) Distribution of experimental states. The red- and blue-marked states indicate the emergence and absence, respectively, of backflow at photon $B$ following an optimal amplitude projective measurement on photon $A$. 
    (b–d)  Transverse probability current $J_x(x, z)$ of photon B for the three red‑marked states in (a). Backflow is characterized by positive probability current for $x < -d/2$ and negative probability current for $x > d/2$. The insets are magnified views of the corresponding regions exhibiting backflow. 
    }
	\label{quantum}
\end{figure}

We further investigate the dependence of the required quantum‑correlation strength for nonlocally manipulating the occurrence of backflow. Werner states of different purities $p$  were prepared on planes at $z = 4134$ mm, $z = 4665$ mm, and $z=5723$ mm, and characterized using concurrence~\cite{2001measurement}, critical radius~\cite{nguyen2019geometry,rong2024sharing}, CHSH parameter~\cite{giustina2015significant}, and probability current. As shown in Fig.~\ref{quantum}, when photon A is projected onto the state
$\left|\theta\right\rangle_A =\cos\theta\left|H\right\rangle - \sin\theta\left|V\right\rangle$ with the optimal parameter $\theta$, the minimal correlation strength required to nonlocally manipulate the emergence of backflow in photon B increases with propagation distance. Specifically, at $z = 4134$ mm, entangled states are sufficient; at $z = 4665$ mm, steerable states become necessary; and at $z = 5723$ mm, only Bell‑nonlocal states can sustain the effect.  
This distance dependence arises from the free-space divergence of the wavepackets, which reduces the local probability density and thereby suppresses the magnitude of the negative current. Instead, stronger quantum correlations are required to generate a negative current capable of overcoming the outward single-slit flow. Experimentally measured concurrence, critical radius, and CHSH parameters, along with the residual probability currents for the states displayed in Fig.~\ref{quantum}(a), are provided in Appendix C.

\section{Conclusion and Discussion}
In summary, we demonstrate nonlocal manipulation of backflow in a double-slit interference system using polarization–path-entangled photons. By combining amplitude and phase manipulation with local measurements on a remote photon, we achieve full manipulation over the emergence, spatial distribution, and propagation dynamics of backflow without directly accessing the interfering system. Weak-measurement reconstruction of probability currents and Bohmian trajectories provides a direct visualization of the manipulation process at single-pixel resolution. Most importantly, using Werner states with tunable correlations, we show that the minimum quantum‑correlation strength required for nonlocally manipulating the emergence of backflow increases with propagation distance, transitioning from entanglement to EPR steering and ultimately to Bell nonlocality.

Our approach enables nonlocal backflow manipulation at the single-photon level, paving the way for promising applications of backflow in optical and quantum information domains. The non-invasive and low-intensity nature, effectively preventing the thermal damage inherent to conventional methods, offers a novel method for nanoparticle trapping, biological imaging, and quantum lithography. Unlike previous experimental studies on backflow, which required the preparation of complex tailored wavefronts \cite{eliezer2020} and relied on measurements using the Shack-Hartmann wavefront sensor \cite{daniel2022,ghosh2023}—both constrained by low spatial resolution—our method achieves significant improvements in simplicity and precision. The generation of backflow here requires only the superposition of two Gaussian wavepackets, achievable in various physical systems such as classical optics \cite{daniel2022}, neutrons \cite{zeilinger1988single}, molecules \cite{cotter2017search}, and electrons \cite{tonomura1989demonstration}. Additionally, the weak-measurement-based backflow detection not only simplifies the experimental setup but also enhances spatial resolution to the level of a single pixel. This advance establishes a powerful and broadly applicable method for the observation and manipulation of backflow.

We conclude with several promising research directions. First, it would be interesting to generalize our results to encompass multi-dimensional motion and multi-particle systems, which could expand the backflow region and uncover novel phenomena. Second, further exploration of the propagation properties and nonlocal characteristics of backflow as photon velocities approach the speed of light could yield deeper insight into its behavior in relativistic regimes. Third, advancing experimental applications of backflow, particularly in improving the precision of particle manipulation and imaging, is crucial, as relevant experimental research is still lacking. Finally, further studies are required to elucidate the fundamental relationship between quantum correlations and backflow.

\appendix   

\section*{\MakeUppercase{Appendix A distance-dependent resource requirement for nonlocal backflow manipulation}}
\addcontentsline{toc}{section}{Appendix A}  
\setcounter{equation}{0}                     
\renewcommand{\theequation}{A\arabic{equation}}

Werner states \cite{werner1989quantum} are a canonical class of mixed states that interpolate between a maximally entangled state and the maximally mixed state (white noise). They play a crucial role in distinguishing between different types of quantum correlations, such as entanglement, Bell nonlocality, and steering. For two qubits, they can be written as
\begin{equation}
\rho_W = p |\Psi^-\rangle \langle\Psi^-| + \frac{1-p}{4} \mathbb{I}_4,
\label{eq:werner_state}
\end{equation}
where $|\Psi^-\rangle =  (|HV\rangle - |VH\rangle)/\sqrt{2}$  is the singlet Bell state, $\mathbb{I}_4$ is the $4 \times 4$ identity matrix,  and $p$ is a purity parameter ($p \in [0, 1]$). 
In this work, we employ Werner states to investigate the minimal correlation strength required to nonlocally manipulate the occurrence of backflow, thereby exploring the connection between backflow and three distinct quantum correlations, and .

Here, we take the amplitude modulation scheme as an example, photon A is projected onto $|m_A\rangle=\cos \theta |H \rangle_A - \sin \theta | V\rangle_A$. Using the identity decomposition $\mathbb{I}_4 = |HH\rangle\langle HH| + |VV\rangle\langle VV| + |VH\rangle\langle VH| + |HV\rangle\langle HV|$, the conditional state of photon B becomes
\begin{equation}
   \rho_{B|m}  =  p |\theta_B\rangle\langle \theta_B| + \frac{1-p}{2} (|H\rangle\langle H| + |V\rangle\langle V|).
\end{equation}
where  $|\theta_B\rangle= \cos \theta |V \rangle_B + \sin \theta | H\rangle_B$. The corresponding probability current for photon B is then given by the mixed-state formula Eq.~\ref{eq:current},
\begin{equation}
J^{\rho_{B|m}}_x(x,z) = p J_x^s(x,z) + \frac{1-p}{2}\left(J_x^H(x,z)+J_x^V(x,z)\right),
\label{eq:j_werner}
\end{equation}
where $J_x^s(x,z)$ is the current associated with the coherent component $|\theta_B\rangle$ (corresponding to the two‑slit superposition), while $J_x^H(x,z)$ and $ J_x^V(x,z)$ are the single‑slit currents associated with the incoherent horizontal and vertical polarization components, respectively.

\begin{figure}[hbt]
    \centering
    \includegraphics[width=0.9\linewidth]{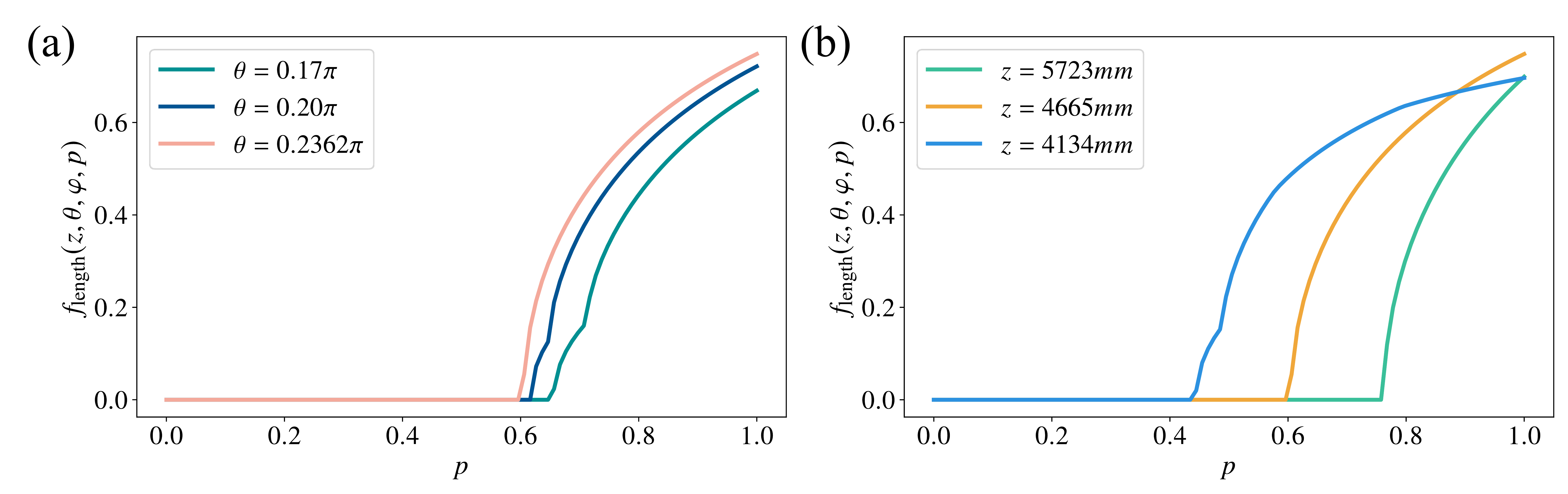}
    \caption{Dependence of backflow interval length $f_{\text{length}}(z,\theta,0,p)$ on state purity $p$. (a) $f_{\text{length}}$ as a function of $p$ for different measurement angles $\theta$ at a fixed distance $z = 4665$ mm.  $\theta = 0.2362 \pi$ corresponds to the optimal angle that minimizes the purity threshold $p_{\mathrm{th}}$ required to nonlocally manipulate the occurrence of backflow. (b) $f_{\text{length}}$ as a function of $p$ at the corresponding optimal angle $\theta_{\mathrm{opt}}(z)$ for different propagation distances $z$. The increase of the minimal threshold $p_{\text{th}}^{\text{min}}(z)$ with $z$ indicates that stronger quantum correlations are required to observe backflow at larger distances.}
    \label{fig:S1}
\end{figure}

As the purity parameter $p$ decreases, the growing weight of the identity term enhances the contributions from the single-slit currents $J_H(x,z)$ and $ J_V(x,z)$. In a double-slit interference, these currents carry a sign opposite to that of any backflow component. Consequently, reducing $p$ suppresses backflow: it prevents it from appearing if initially absent and reduces its magnitude if already present.

To characterize this behavior quantitatively, we define a function $f_{\text{length}}(z,\theta,\varphi,p)$ that quantifies the total spatial extent of backflow along the $x$-axis. 
In our numerical analysis, the probability current $J_x^{\rho_{B|m}}(x,z)$ is evaluated within the transverse window $x \in [-4, 4]$, which encloses all observed backflow regions. For computational convenience, the continuous integral is converted into a discrete sum over all contiguous backflow intervals identified inside this window.
Specifically, the window is discretized into $5000$ uniformly spaced points $x_i = -4 + 8(i-1)/4999$ ($i=1,\dots,5000$), yielding a spatial resolution of $\Delta x \approx 0.0016$, and $J_x^{\rho_{B|m}}(x_i,z_i)$ is computed at each point for a given $z_i$.
In the double‑slit interference, backflow occurs where $J_x^{\rho_{B|m}}(x,z)>0$ for $x < -d/2$  and $J_x^{\rho_{B|m}}(x,z) < 0$ for $x > d/2$ (here $d=3$ mm). The total backflow length is then obtained by summing the lengths of all contiguous intervals that satisfy these sign conditions; interval boundaries are refined using linear interpolation at each sign change of $J_x(x_i,z_i)$.  Mathematically,
\begin{equation}
f_{\text{length}}(z,\theta,\varphi,p) = L_{\text{right}} + L_{\text{left}},
\label{eq:total_backflow}
\end{equation}
where
\begin{align}
L_{\text{right}} = \sum_{k=1}^{N_r} \Bigl[ \tilde{x}_{k,\text{right}}^{(e)} - \max\bigl(\frac{d}{2},\; \tilde{x}_{k,\text{right}}^{(s)}\bigr) \Bigr], \quad
L_{\text{left}} = \sum_{l=1}^{N_l} \Bigl[ \min\bigl(-\frac{d}{2},\; \tilde{x}_{l,\text{left}}^{(s)}\bigr) - \tilde{x}_{l,\text{left}}^{(e)} \Bigr],
\label{eq:left_backflow}
\end{align}
with $N_r$ and $N_l$ denoting the number of continuous backflow intervals in the right  ($x > d/2$) and  left ($x < -d/2$) regions, respectively. Here, $\tilde{x}_{k,\text{right}}^{(s)}$ and $\tilde{x}_{k,\text{right}}^{(e)}$ represent the starting and ending points of the $k$-th continuous backflow interval on the right side. These points correspond to the positions where the current changes sign: $\tilde{x}_{k,\text{right}}^{(s)}$ is where the current turns from non‑negative to negative, and $\tilde{x}_{k,\text{right}}^{(e)}$ from negative back to non‑negative. Similarly, $\tilde{x}_{l,\text{left}}^{(s)}$ and $\tilde{x}_{l,\text{left}}^{(e)}$ denote the boundaries of the $l$-th backflow interval on the left side, with the sign changes reversed due to the opposite flow direction. The refined boundary positions $\tilde{x}^{(s)}$ and $\tilde{x}^{(e)}$ are obtained by linear interpolation.
\begin{equation}
\tilde{x}_{\text{boundary}} = x_0 + \frac{0-J_x^{\rho_{B|m}}(x_0)}{J_x^{\rho_{B|m}}(x_1)-J_x^{\rho_{B|m}}(x_0)}(x_1-x_0),
\label{eq:boundary_interpolation}
\end{equation}
where $x_0$ and $x_1$ are adjacent grid points satisfying $J_x^{\rho_{B|m}}(x_0)J_x^{\rho_{B|m}}(x_1)<0$. 

By analyzing the dependence of $f_{\text{length}}(z,\theta,\varphi,p)$ on $p$ for fixed $(z, \theta, \varphi)$, we find that the backflow length is monotonically non-decreasing with $p$ in the far feild (see Figure.~\ref{fig:S1}). Hence, for any given parameter setting $(z, \theta, \varphi)$, there exists a unique purity threshold $p_{\text{th}}$ such that backflow occurs only for $p > p_{\text{th}}$. 
To obtain the minimal purity required for nonlocally manipulate the occurrence of backflow at a given propagation plane $z$, we optimize over the measurement angle $\theta$ on photon A.  Scanning $\theta$ from $0$ to $\pi/2$ yields a threshold $p_{\text{th}}(\theta)$ for each angle. The global minimum, $p_{\text{th}}^{\text{min}}(z)$, defines the minimal purity threshold for that plane, corresponding to the optimal angle $\theta_{opt}$. Figure~\ref{fig:S1}(a) shows $f_{\text{length}}(4655\mathrm{mm},\theta,0,p)$ versus $p$ for three measurement angles; the optimal angle minimizing the threshold is $\theta = 0.2362\pi$. 
 
Figure~\ref{fig:S1}(b) further displays $f_{\text{length}}(z,\theta_{opt},0,p)$ as a function of $p$ for three different propagation distances $z$, each evaluated at its respective optimal angle $\theta_{\text{pot}}(z)$ for each distance (with $\theta_{\mathrm{opt}}(4134\ \text{mm}) = 0.2123 \pi$ and $\theta_{\mathrm{opt}} (5723\ \text{mm})= 0.2387\pi$). The results show that $p_{\text{th}}^{\text{min}}(z)$ increases with $z$ in the far field, indicating that the strength of quantum correlations required to observe backflow grows with propagation distance, progressing from entanglement to steering, and finally to Bell nonlocality.  
It should be noted that all results from the amplitude-modulation scheme can be similarly obtained in the phase-modulation scheme.

\section*{\MakeUppercase{Appendix B Additional experimental Backflow Distributions under Static Phase Modulation}}
\addcontentsline{toc}{section}{Appendix B}
\setcounter{equation}{0}
\renewcommand{\theequation}{B\arabic{equation}}

\begin{figure}[hbt]
    \centering
    \includegraphics[width=0.85\linewidth]{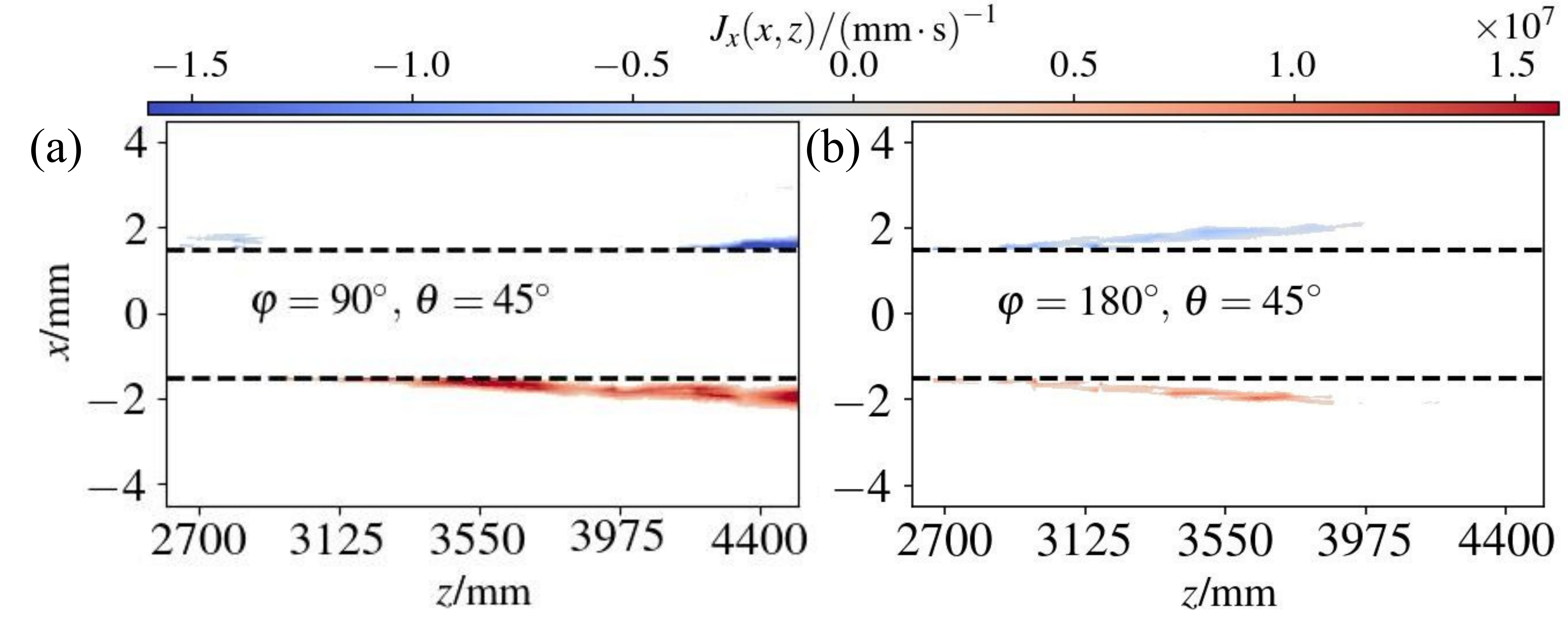}
    \caption{Manipulation of backflow via static phase modulation. Backflow regions of photon B  when photon A is projected onto $(\left|H\right\rangle - e^{-i\varphi} \left|V\right\rangle)/\sqrt{2}$ for $\varphi =  90^\circ$ and $180^\circ$. }
    \label{fig:S0}
\end{figure}

In the static phase modulation scheme,  photon A is post-selected in the polarization state $|m_A\rangle= (\left|H\right\rangle - e^{-i\varphi(z)} \left|V\right\rangle)/\sqrt{2}$, where the relative phase remains constant during propagation ($\varphi(z) = \varphi$). Figure~\ref{fig:S0} displays the spatial distributions of the backflow regions for $\varphi =  90^\circ$ and $180^\circ$. Clearly, the backflow regions are symmetric about $x = 0$ when $\varphi= 0^\circ$ and $180^\circ$, whereas the symmetry is broken for $\varphi = 90^\circ$ and $270^\circ$. 
Notably, the spatial profile and intensity of the backflow near the upper slit at $\varphi = 90^\circ$ closely match those near the lower slit at $\varphi = 270^\circ$.

\section*{\MakeUppercase{Appendix C Additional experimental results on Distance-Dependent Resource Requirement for Backflow Manipulation}}
\addcontentsline{toc}{section}{Appendix C}
\setcounter{equation}{0}
\renewcommand{\theequation}{C\arabic{equation}}

\begin{figure}[ht!]
    \centering
    \includegraphics[width=0.8\linewidth]{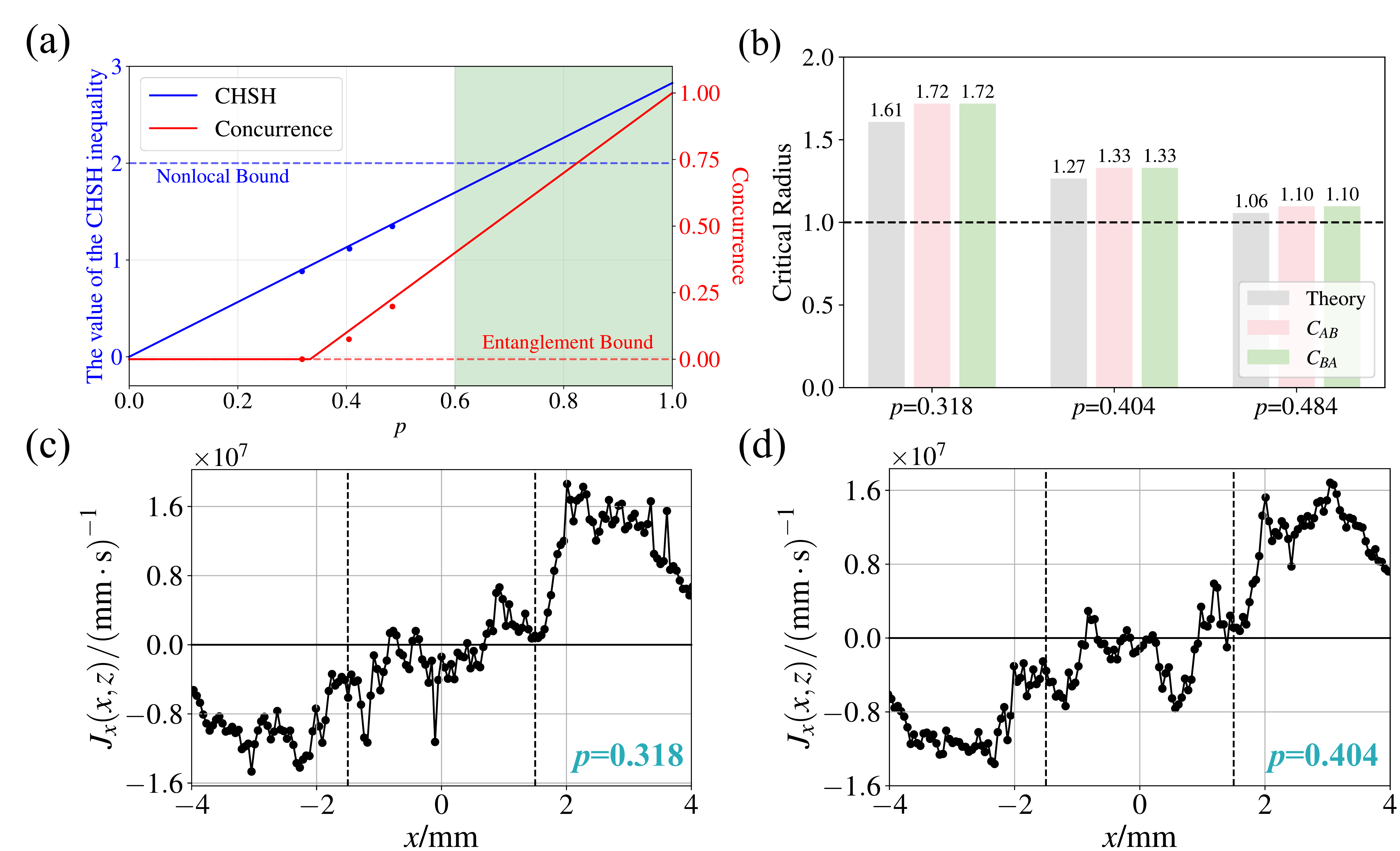}
    \caption{Experimental results at the propagation distance $z = 4134\,\text{mm}$. (a) Measured CHSH parameter (left axis) and Concurrence (right axis) of Werner states with three different purities $p_1^{z_1}=0.318$, $p_2^{z_1}=0.404$, and $p_3^{z_1}=0.484$. 
(b) Theoretical (gray) and experimentally obtained (red and green) critical radius for the states shown in (a). (c, d) Transverse probability current $J_x(x, z)$ of photon B following an optimal amplitude projective measurement on photon $A$, for Werner states with purity $p_1^{z_1}$ (c) and $p_2^{z_1}$ (d).}
    \label{fig:S2}
\end{figure}
\begin{figure}[hb!]
    \centering
    \includegraphics[width=0.8\linewidth]{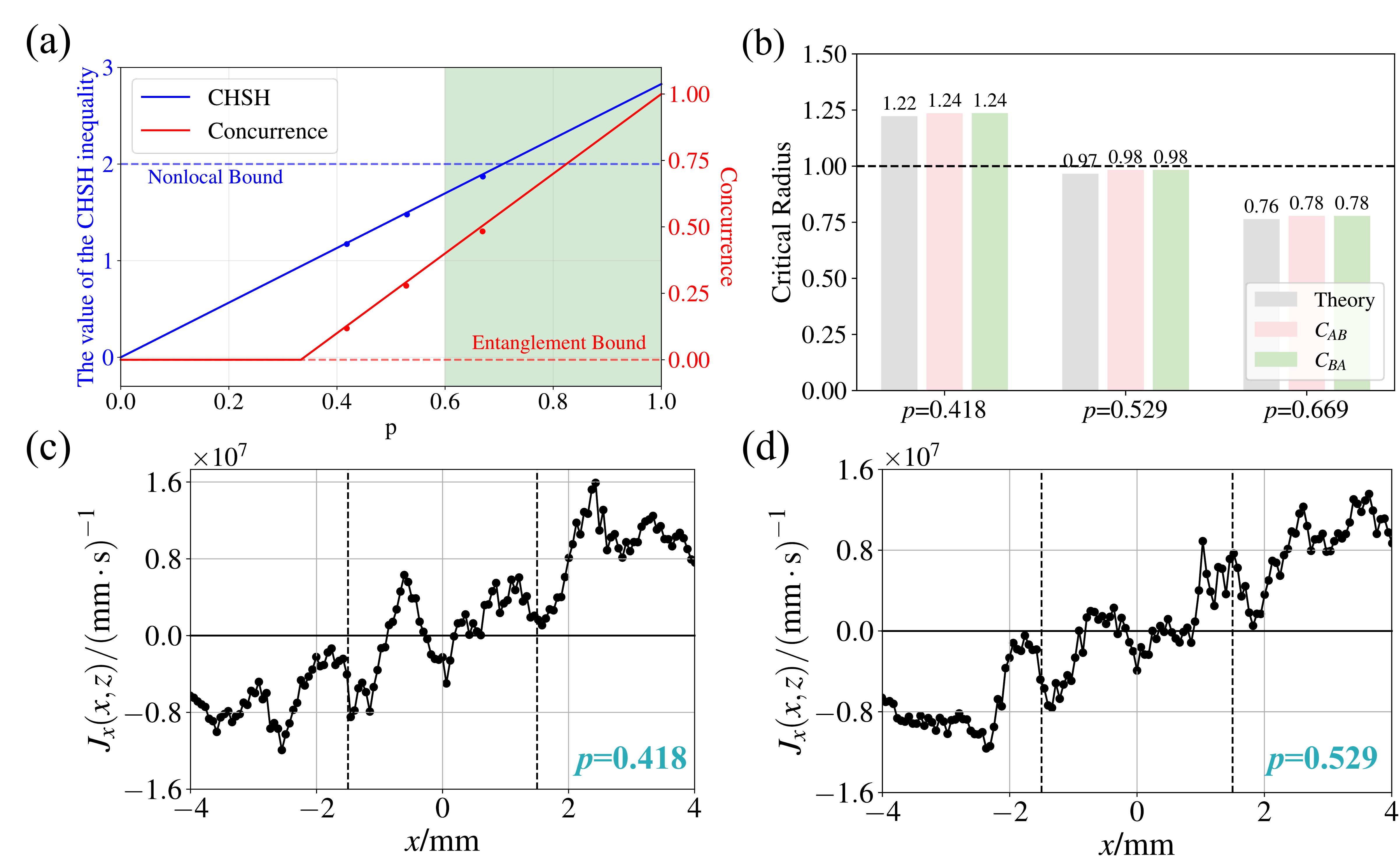}
    \caption{Experimental results at the propagation distance $z = 4665\,\text{mm}$.(a) Measured CHSH parameter (left axis) and Concurrence (right axis) of Werner states with three different purities $p_1^{z_2}=0.418$, $p_2^{z_2}=0.529$, and $p_3^{z_2}=0.669$.
 (b) Theoretical (gray) and experimentally obtained (red and green) critical radius for the states shown in (a).
(c, d) Transverse probability current $J_x(x, z)$ of photon B following an optimal amplitude projective measurement on photon $A$, for Werner states with purity $p_1^{z_2}$ (c) and $p_2^{z_2}$ (d).}
    \label{fig:S3}
\end{figure}

To investigate the minimal correlation strength required to nonlocally manipulate the occurrence of backflow in the far field, we prepared Werner states with three purities, labeled $p^{z_i}_1, p^{z_i}_2, p^{z_i}_3$, at three distinct propagation distances $z=4134\,\text{mm}$, $z=4665\,\text{mm}$, and $z=5723\,\text{mm}$.
The type of correlation for each state is quantified by  concurrence for entanglement~\cite{2001measurement}, critical radius for steering~\cite{nguyen2019geometry, rong2024sharing}, and CHSH parameter for Bell nonlocality~\cite{giustina2015significant}. State analysis is performed with the polarization‑analysis module shown in Fig.~2(b) and (d) of the main text, consisting of a quarter‑wave pblate, a half‑wave plate, and a polarization beam splitter. In each case, Alice implements the amplitude‑modulation measurement $|m_A\rangle=\cos \theta |H \rangle - \sin \theta | V\rangle$  with the optimal angle $ \theta_{opt}$ chosen as described in Appendix B.  The transverse probability‑current distribution $J_x(x, z)$ of photon B is then reconstructed via the backflow‑analysis module shown in Fig.~2(c). 

\begin{figure}
    \centering
    \includegraphics[width=0.8\linewidth]{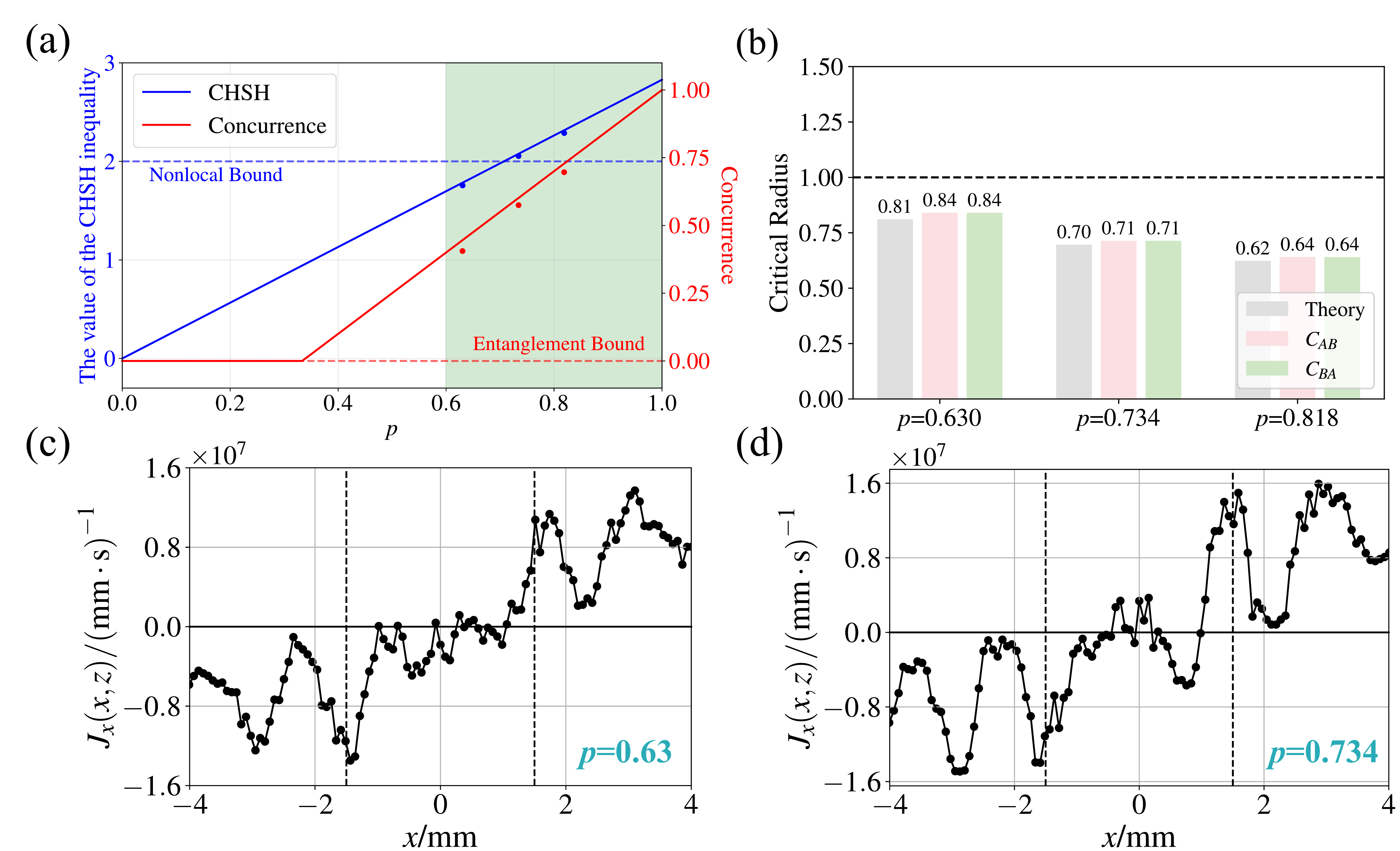}
    \caption{Experimental results at the propagation distance $z = 5723\, \text{mm}$.(a) Measured CHSH parameter (left axis) and Concurrence (right axis) of Werner states with three different purities $p_1^{z_3}=0.630$, $p_2^{z_3}=0.734$, and $p_3^{z_3}=0.818$.
(b) Theoretical (gray) and experimentally obtained (red and green) critical radius for the states shown in (a). (c, d)      Transverse probability current $J_x(x, z)$ of photon B following an optimal amplitude projective measurement on photon $A$, for Werner states with purity $p_1^{z_3}$ (c) and $p_2^{z_3}$ (d).}
    \label{fig:S4}
\end{figure}

Experimental results in Figs.~\ref{fig:S2}--\ref{fig:S4} show that a measurement on photon A can induce backflow in photon B only when the shared Werner state possesses sufficiently strong quantum correlations. At \( z = 4134\,\text{mm} \), backflow occurs exclusively for the strongly entangled state (\( p^{z_1}_3 \)) [main text, Fig.~5(b)], whereas states with no (\( p^{z_1}_1 \)) or weak (\( p^{z_1}_2 \)) entanglement display no backflow  [Figs.~\ref{fig:S2}(c,d)]. At the larger distance $z=4665\,\text{mm}$, backflow appears only for the strongly steerable state (\( p^{z_2}_3 \)) [main text, Fig.~5(c)]; no backflow is observed for states with no (\( p^{z_2}_1 \)) or weak (\( p^{z_2}_2 \)) steerability [Figs.~\ref{fig:S3}(c,d)]. When the propagation distance further increased to  $z=5723$,\text{mm}, a clear backflow signal is observed solely for the state exhibiting strong Bell nonlocality (\( p^{z_3}_3 \)) [main text, Fig.~5(d)], whereas states with no (\( p^{z_3}_1 \)) or weak (\( p^{z_3}_2 \)) Bell nonlocality show none[Figs.~\ref{fig:S4}(c,d)].

These results reveal that inducing backflow at a given propagation distance requires a minimum strength of a specific quantum correlation. This required strength increases with propagation distance---progressing from entanglement to steering to Bell nonlocality---in direct agreement with the theoretically predicted rise of the minimal purity threshold \( p_{\mathrm{th}}^{\min}(z) \).

\begin{backmatter}
\bmsection{Funding}
This work was supported by the National Key Research and Development Program of China (Grant No. 2025YFE0217700), the Shandong Provincial Natural Science Foundation (Grants No. ZR2024LLZ003 and ZR2026LLZ016), the Fundamental Research Funds for the Central Universities (Grants No. 202364008). 
	
\bmsection{Disclosures}
The authors declare no conflicts of interest.

\bmsection{Data availability} The data sets generated during and/or analyzed during the current study are available from the corresponding author on reasonable request.

\end{backmatter}


\bibliography{sample}






\end{document}